\documentclass[conference]{IEEEtran}
\IEEEoverridecommandlockouts
\usepackage{cite}
\usepackage{amsmath,amssymb,amsfonts}
\usepackage{graphicx}
\usepackage{textcomp}
\usepackage{xcolor}
\usepackage{amsthm}
\usepackage[english]{babel}
\usepackage[font=small,labelfont=bf]{caption}
\usepackage{caption}
\usepackage{graphicx}
\usepackage{subcaption}
\usepackage{multirow}
\usepackage{floatrow}
\usepackage{algorithm}% http://ctan.org/pkg/algorithms
\usepackage{algpseudocode}% http://ctan.org/pkg/algorithmicx
\usepackage{subcaption}
\usepackage{mwe}
\usepackage{orcidlink}
\usepackage{array}
\usepackage{hyperref}
\def\BibTeX{{\rm B\kern-.05em{\sc i\kern-.025em b}\kern-.08em
    T\kern-.1667em\lower.7ex\hbox{E}\kern-.125emX}}
\newtheorem{definition}{Definition}
\newtheoremstyle{definition} % name
    {\topsep}                    % Space above
    {\topsep}                    % Space below
    {\itshape}                   % Body font
    {}                           % Indent amount
    {\scshape}                   % Theorem head font
    {.}                          % Punctuation after theorem head
    {.5em}                       % Space after theorem head
    {}  % Theorem head spec (can be left empty, meaning ‘normal’)
\makeatletter
\newcommand{\linebreakand}{%
  \end{@IEEEauthorhalign}
  \hfill\mbox{}\par
  \mbox{}\hfill\begin{@IEEEauthorhalign}
}
\makeatother % changes the catcode of @ back to 12

\addto\captionsenglish{\renewcommand{\figurename}{Fig.}}
\addto\extrasenglish{%
}

\begin{document}
\title{Defining Cases and Variants for Object-Centric Event Data\\
\thanks{The authors would like to thank the Marga und Walter Boll-Stiftung for 
the kind support within the research project.
}
}

\author{
\iffalse \IEEEauthorblockN{Jan Niklas Adams}
\IEEEauthorblockA{\textit{Process and Data Science} \\
\textit{RWTH Aachen University}\\
Aachen, Germany \\
niklas.adams@pads.rwth-aachen.de}
\and
\IEEEauthorblockN{Daniel Schuster}
\IEEEauthorblockA{\textit{Institute for Applied Information Technology} \\
\textit{Fraunhofer}\\
Sankt Augusting, Germany\\
schuster@fit.fraunhofer.de}\fi
\iffalse
\IEEEauthorblockN{Jan Niklas Adams,
Daniel Schuster}
\IEEEauthorblockA{\textit{Process and Data Science}\\
\textit{RWTH Aachen University}\\
Aachen, Germany\\
\{niklas.adams,schuster\}@pads.rwth-aachen.de}\fi

\IEEEauthorblockN{Jan Niklas Adams\orcidlink{0000-0001-8954-4925}}
\IEEEauthorblockA{\textit{Chair for Process and Data Science (PADS)} \\
\textit{RWTH Aachen University},
Aachen, Germany \\
niklas.adams@pads.rwth-aachen.de}
\and
\IEEEauthorblockN{Daniel Schuster\orcidlink{0000-0002-6512-9580}}
\IEEEauthorblockA{\textit{Institute for Applied Information Technology (FIT)} \\
\textit{Fraunhofer},
Sankt Augustin, Germany\\
daniel.schuster@fit.fraunhofer.de}
\linebreakand

\IEEEauthorblockN{Seth Schmitz\orcidlink{0000-0002-0179-0759},
Günther Schuh\orcidlink{0000-0002-6076-0701}}
\IEEEauthorblockA{\textit{Laboratory for Machine Tools (WZL)}\\
\textit{RWTH Aachen University},
Aachen, Germany\\
\{s.schmitz,g.schuh\}@wzl.rwth-aachen.de}
\and
%\linebreakand
\IEEEauthorblockN{Wil M.P. van der Aalst\orcidlink{0000-0002-0955-6940}}
\IEEEauthorblockA{\textit{Chair for Process and Data Science (PADS)} \\
\textit{RWTH Aachen University},
Aachen, Germany \\
wvdaalst@pads.rwth-aachen.de}
}

\maketitle

\begin{abstract}

The execution of processes leaves traces of event data in information systems. These event data can be analyzed through process mining techniques. For traditional process mining techniques, one has to associate each event with exactly one object, e.g., the company's customer. Events related to one object form an event sequence called a case. A case describes an end-to-end run through a process. The cases contained in event data can be used to discover a process model, detect frequent bottlenecks, or learn predictive models. However, events encountered in real-life information systems, e.g., ERP systems, can often be associated with multiple objects. The traditional sequential case concept falls short of these so-called object-centric event data since these data exhibit a graph structure. One might force object-centric event data into the traditional case concept by flattening it. However, flattening manipulates the data and removes information. Therefore, a concept analogous to the case concept of traditional event logs is necessary to enable the application of different process mining tasks on object-centric event data. In this paper, we introduce the case concept for object-centric process mining: process executions. These are graph-based generalizations of cases as considered in traditional process mining. Furthermore, we provide techniques to extract process executions. Based on these executions, we determine equivalent process behavior with respect to an attribute using graph isomorphism. Equivalent process executions with respect to the event's activity are object-centric variants, i.e., a generalization of variants in traditional process mining. We provide a visualization technique for object-centric variants. The contribution's scalability and efficiency are extensively evaluated. Furthermore, we provide a case study showing the most frequent object-centric variants of a real-life event log. Our contributions might be used as a basis to adapt traditional process mining techniques by researchers and to generate initial control-flow insights into object-centric event logs by practitioners.
\end{abstract}


\section{Introduction}
\textit{Process mining}~\cite{ProcessMiningDSIA} is an umbrella term for techniques discovering knowledge about processes from \textit{event data} that these processes generated. These event data come as an \textit{event log}. Each event of an event log describes an activity executed in the process, together with its associated data.

%One of the most fundamental concepts in process mining is the one of cases. (Explain)
One of the most fundamental concepts in process mining are \textit{cases}. A case is an event sequence describing one observed end-to-end behavior in a process. Each event is associated to the \textit{object} for which the event was conducted, e.g., the customer. This object is called the case of the event and each object can be associated to multiple events, forming the event sequence of end-to-end behavior. The event sequences of all objects are the fundamental starting point of many process mining algorithms: control-flow visualization~\cite{Cortado}, i.e., frequent sequences of conducted actions, bottleneck analysis~\cite{bottleneck_queuing}, or outcome prediction~\cite{LSTMPrediction}. An event log in traditional process mining exhibits the structure depicted in \autoref{fig:sequence1}: a collection of event sequences, one for each case.
%For what are cases used? (Visualizing control flow variations, bottleneck analysis, etc)

{\setlength{\belowcaptionskip}{-8pt}
\begin{figure}[t]
  \centering
  \subfloat[Traditional event logs.]{\includegraphics[width=0.49\columnwidth]{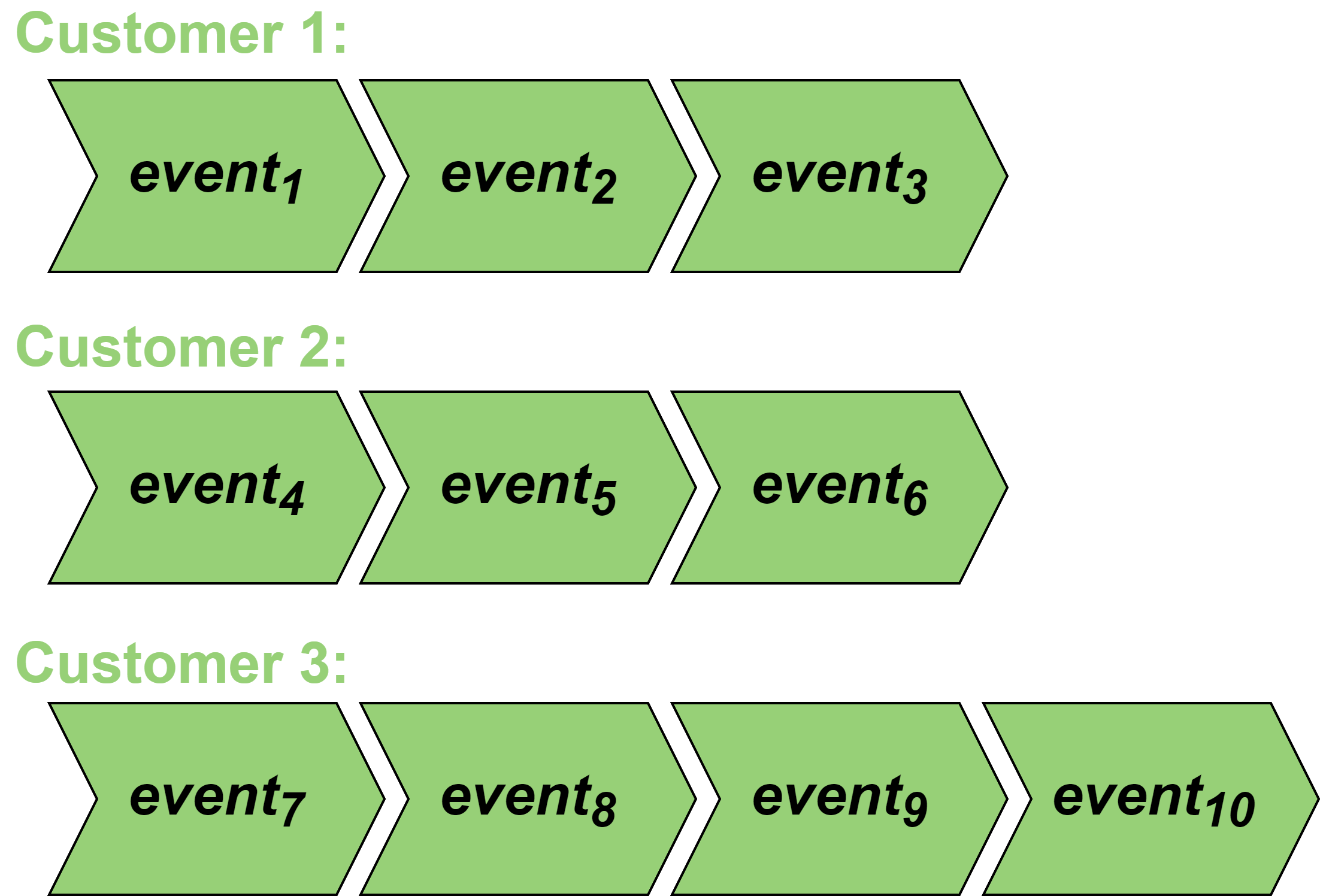}\label{fig:sequence1}}
  \hfill
  \subfloat[Object-centric event logs.]{\includegraphics[width=0.49\columnwidth]{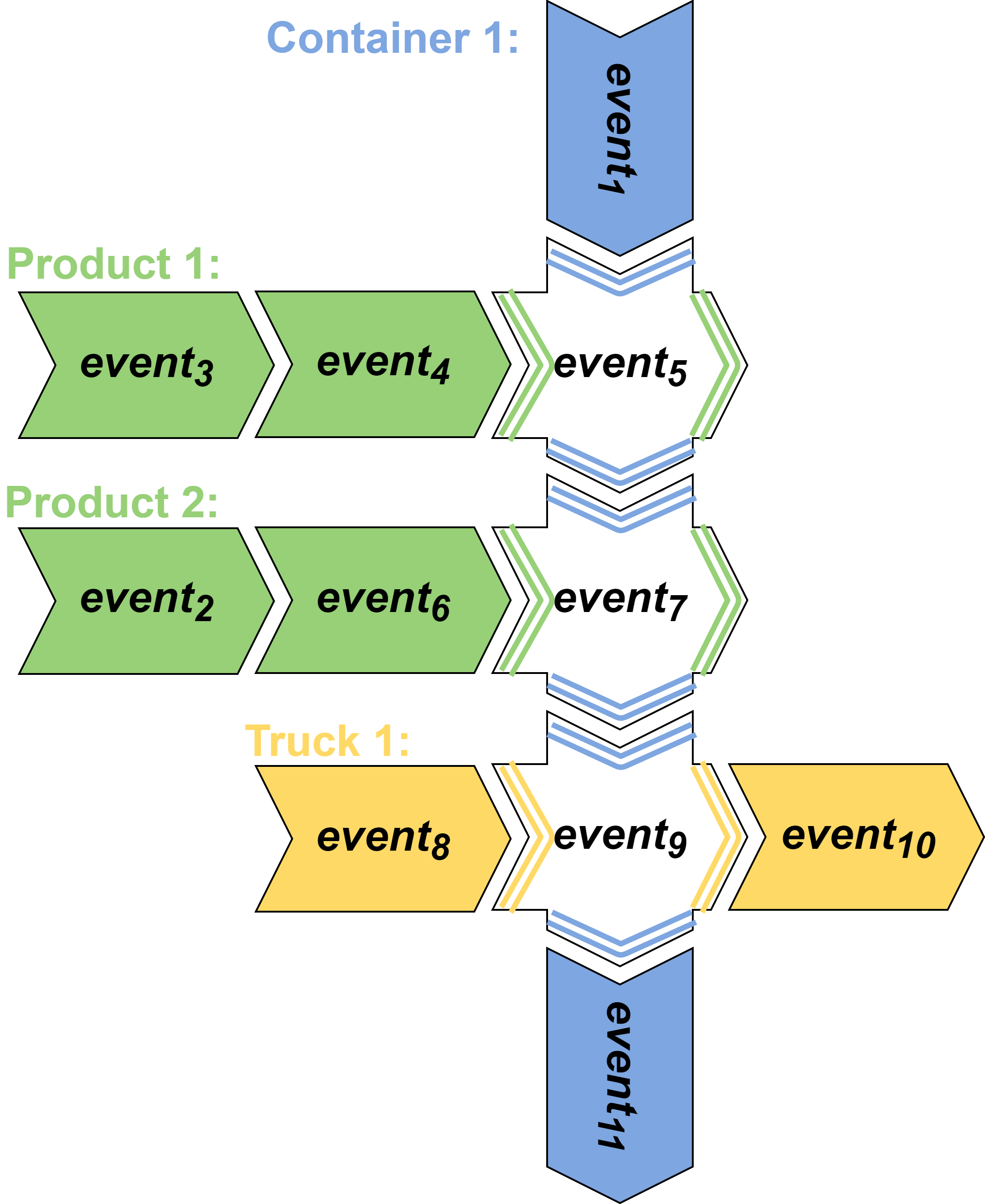}\label{fig:sequence2}}
  \caption{Traditional event logs take the form of (a): Each event is associated to exactly one object, called the case. Object-centric event logs (b) drop this restriction, events can be associated to multiple objects of different types.}
  \label{fig:sequences}
\end{figure}}

%OC data descirbes event data where one event is related to obejcts of multiple types
\textit{Object-centric event data}~\cite{OCPN} is a generalized form of traditional event data. We drop the assumption that one event can only be related to one object.  Each object is still associated to a sequence of events, but one event may belong to multiple sequences. An example of the resulting structure of object-centric event data is depicted in \autoref{fig:sequence2}. Some events are shared between objects, e.g., $\mathit{event}_5$. The sequential structure of the event data is lost by dropping the assumption of single-object association. \emph{The event data takes a graph structure.} Furthermore, objects are associated to different \textit{object types}. An information system may record data for the conducted actions of objects of different types, e.g., products and delivery objects in an ordering process. In traditional process mining, all objects are assumed to be of the same type, e.g., all events describe an action conducted for a customer, leading to homogeneous events. In object-centric event logs, the homogeneity assumption is dropped, leading to differently typed objects and events. In conclusion, traditional event data describe homogeneously typed event sequences, while object-centric event data describe heterogeneously typed event graphs.
Object-centric event data is closer to the reality experienced in many real-life information systems: It is encountered in production processes~\cite{DataModelEndToEnd}, high volume manufacturing data~\cite{BigDataProductionOverview}, and order-to-cash processes~\cite{CausalProcessMining}. Object-centric event data exhibiting a graph structure was already observed in~\cite{GraphDatabases,PrecisionFitnessOCPM}

%A sequential case concept is insufficient to represent object-centric event data.
Most process mining techniques rely on the existence of cases. To apply these techniques to object-centric event logs, the case concept and object-centric event logs must be connected. There are two ways to bridge the gap between the case concept and object-centric event logs: Either moving object-centric event logs to the traditional case concept or moving the traditional case concept to object-centric event logs. First, one can force object-centric event data into traditional event log format, enforcing homogeneity and sequentiality. This is called \textit{flattening}~\cite{OCPMDivergenceAndconvergence}. Second, one can generalize the concept of cases from homogeneous sequences to heterogeneous graphs and adapt process mining techniques accordingly. When flattening an object-centric event log, one chooses an object type (also: case notion), removes events with no objects of this type, and duplicates events with multiple objects of that type. The output of flattening is a traditional event log. While flattening is fast and straightforward, it manipulates the event data: information about diversity in object types is discarded and a sequential structure removes dependencies contained in the event data~\cite{CausalProcessMining,OCPMDivergenceAndconvergence}. For these reasons, we aim to move the case concept towards object-centric event data in this paper. We provide the following contributions:
\begin{itemize}
    \item[\textbf{C1}] We generalize the case concept of traditional process mining from homogeneous sequences to heterogeneous graphs for object-centric event logs. These are called process executions.
    \item[\textbf{C2}] We provide a general approach for extracting process executions from an object-centric event log. Furthermore, we provide two specific extraction techniques.
    \item[\textbf{C3}] We use graph isomorphism to determine equivalent process executions. Isomorphic graphs can be used to group equivalent behavior, e.g., object-centric variants for equivalent control-flow behavior.
    \item[\textbf{C4}] We propose an algorithm to visualize equivalent behavior with a focus on visualizing object-centric variants. This visualization is an extension of traditional variant visualization.
\end{itemize}
These contributions aim to provide a foundation for moving process mining from traditional event data to object-centric event data. Using process executions, one may adapt existing algorithms and create new algorithms to discover new insights while leveraging the full, available information. 

We discuss related work for this paper in \autoref{sec:relwork} and continue with basic definitions on object-centric event data in \autoref{sec:prelim}. These definitions build the basis for the introduction of process executions and their extraction in \autoref{sec:execandequiv}. We discuss equivalent process executions and variant visualization in \autoref{sec:variants}. The technical side of the contributions, i.e., scalability and efficiency, are evaluated in \autoref{sec:evaluation}. We demonstrate the utility of our contributions in a case study in \autoref{sec:case}. We conclude the paper in \autoref{sec:conclusion} and provide directions for future contributions.

\section{Related Work}
\label{sec:relwork}
{\setlength{\belowcaptionskip}{-10pt}
\begin{figure}[t]
    \centering
    \includegraphics[width=0.7\columnwidth]{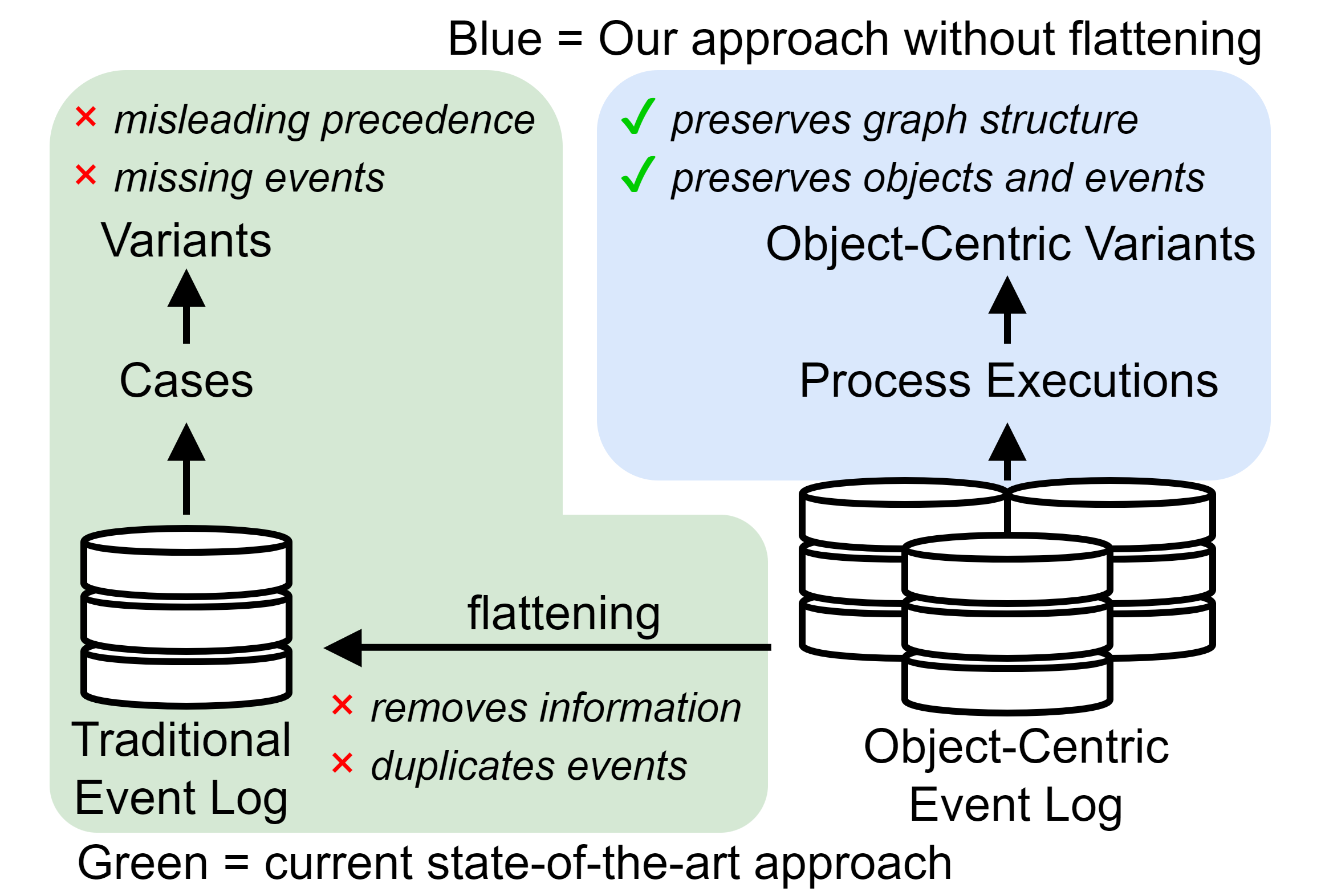}
    \caption{Conceptual difference between our work and the current state-of-the-art for process mining on object-centric event logs. Current approaches flatten the event log leading to manipulated and removed information. The generated results might, therefore, also be misleading. We propose a graph-based case concept for object-centric event data called process executions. These can accurate represent object-centric event data.}
    \label{fig:relwork}
\end{figure}}
Object-centric process mining deals with generating insights for event data with multiple objects. The problem of object multiplicity in real-life information systems was already formulated several years ago~\cite{manifesto}. Early approaches dealt with the problem from a modeling perspective~\cite{ProcletManytoMany,ArtifactsDiscovery} and with different object types being investigated separately. Object-centric process mining introduced a data-driven way to approach the problem and has recently gained attention~\cite{OCPN,CausalProcessMining,GraphDatabases,OCPredictive,OPERA}. So far, different tasks of process mining have been adapted to the object-centric setting. We group these into three categories corresponding to this paper's focus: Approaches operating case agnostically, approaches working with flattening~\cite{OCPMDivergenceAndconvergence} and approaches working with graphs as case concept. The techniques that were introduced for discovery~\cite{OCPN} and conformance checking~\cite{PrecisionFitnessOCPM} work without the explicit use of a case concept. Galanti et al.~\cite{OCPredictive} propose to flatten the event log to apply predictive process mining techniques to the flattened event log. The flattened event data is a traditional event log and can be used as input to all traditional process mining techniques. However, flattening is associated to the problems of \textit{deficiency}, \textit{convergence}. and \textit{divergence}~\cite{OCPMDivergenceAndconvergence}. By not flattening the data and generalizing the case concept to a graph, these problems can be avoided. The conceptual differentiation between our work and previous work based on flattening is depicted in \autoref{fig:relwork}. To the best of our knowledge, we are the first paper defining a graph-based case concept and variants for object-centric event data. 
\section{Preliminaries and Event Data}
\label{sec:prelim}
Given a set $X$, a sequence $\sigma \in X^*$ of length $n\in \mathbb{N}$ assigns an enumeration to elements of the set $\sigma:\{1,\ldots,n\}\rightarrow X$. We use the notation $\sigma {=} \langle \sigma_1,\ldots, \sigma_n \rangle$. For an element $x\in X$ and a sequence $\sigma \in X^*$, we overload the notation $x\in\sigma $, expressing the occurrence of element $x$ in the sequence $x\in range(\sigma)$.

We are dealing with event data of different object types. $\mathcal{T}$ defines the universe of types. There can be multiple instantiations of one type. We refer to each instantiation as an object. $\mathcal{O}$ defines the universe of objects. Each object is of one type $\pi_{type}:\mathcal{O}\rightarrow \mathcal{T}$. We define an event log containing events and objects of different types, assigning each object to an event sequence. $\mathcal{E}$ denotes the universe of event identifiers, $\mathcal{A}$ denotes the universe of event attributes and $\mathcal{V}$ denotes the universe of attribute values.
\begin{definition}[Event Log]
An event log $L{=}(E,O,\allowbreak \mathit{trace},\allowbreak \mathit{attr})$ is a tuple where 
\begin{itemize}
    \item $E\subseteq \mathcal{E}$ is a set of events,
    \item $O\subseteq \mathcal{O}$ is a set of objects,
    \item $\mathit{trace} : O {\rightarrow} E^*$ maps each object to an event sequence,
    \item $attr : E {\times} \mathcal{A} {\nrightarrow} \mathcal{V}$ maps event attributes onto values.
\end{itemize}
\iffalse
We denote the event order relations for all sequences of an event log $L$ by $\prec_L = \{(e,e')\in E{\times} E\mid \allowbreak \exists_{o{\in}O}\ trace(o){=} \langle e_1,\ldots ,e_n\rangle \allowbreak \exists_{1\leq i< j \leq n} \allowbreak (e{=}e_i \wedge e'{=}e_j)\}$. Order relations must be consistent among all sequences, i.e., $(e,e')\in \prec_L \Rightarrow (e',e) \notin \prec_L$.\fi
\iffalse
The transitive closure of $\prec_L = \{(e,e')\in E{\times} E\mid \allowbreak \exists_{o{\in}O}\ trace(o)= \langle e_1,\ldots ,e_n\rangle \allowbreak \exists_{1\leq i< j \leq n} \allowbreak (e=e_i \wedge e'=e_j)\}$ is a strict partial order\footnote{Irreflexivity, transitivity and asymmetry must be fulfilled.}. We call its transitive reduction the event-object graph~\cite{PrecisionFitnessOCPM}.\fi
\end{definition}
We define some further notations used throughout the paper.
\begin{definition}[Further Notations]
Let $L{=}(E,O,\allowbreak \mathit{trace},\mathit{attr})$ be an event log. We define the following notations:
\begin{itemize}
    \item $obj_L(e) = \{o \in O \mid e\in \mathit{trace}(o)\}$ for $e \in E$ denote the objects associated to an event.
    \item $con_L = \{(e,e')\in E\times E \mid \exists_{o{\in} O} \ \mathit{trace}(o) = \langle e_1,\ldots, e_n\rangle\allowbreak \exists_{1{\leq}i{<}n} \ e=e_i \wedge e'=e_{i+1}\}$ defines the directly-follows relationships for all events and all objects.
\end{itemize}
\end{definition}
An example of an event log is tabularly depicted in \autoref{fig:event_data_example}. Multiple objects are given: o1, o2, m1, m2, m3, and m4. Objects are of different types, e.g., o1 is of type Type1. Each object is associated to a sequence of events, e.g., $trace(o1)=\langle e_3,e_4,e_6\rangle$. 
\iffalse Two events should not appear in conflicting order in two different event sequences.\fi
\iffalse
We define further notations based on the event log.
\iffalse\begin{definition}
Given an event log $L=(E,O,trace,attr)$ we define the following notations:
\begin{itemize}
    \item $obj_L(e) = \{o \in O \mid e\in trace(o)\}$ for $e\in E$
    \item $con_L = \{(e,e')\in E\times E \mid \exists_{o{\in} O} \ trace(o) = \langle e_1,\ldots, e_n\rangle\allowbreak \exists_{1{\leq}i{<}n} \ e=e_i \wedge e'=e_{i+1}\}$
   \iffalse
    \item $\leq_L = \{(e,e') \in E\times E \mid \exists_{o{\in}O}trace(o) = \langle e_1,\ldots, e_n\rangle\allowbreak \exists_{1{\leq}i{\leq}j{\leq}n} e= e_i \wedge e'=e_j\}$\fi
\end{itemize}
\end{definition}\fi
The notation $obj_L(e) = \{o \in O \mid e\in trace(o)\}$ for $e\in E$ defines the objects associated to an event $e$, i.e., all objects that contain $e$ within their event sequence. $con_L = \{(e,e')\in E\times E \mid \exists_{o{\in} O} \ trace(o) = \langle e_1,\ldots, e_n\rangle\allowbreak \exists_{1{\leq}i{<}n} \ e=e_i \wedge e'=e_{i+1}\}$ contains all relationships of events that directly follow each other in any of the event sequences. \fi

The right-hand side of \autoref{fig:event_data_example} depicts the events enriched with the information of $\mathit{obj}_L$ and $\mathit{con}_L$. The result is a graph (also: event-object graph~\cite{PrecisionFitnessOCPM}), where the events form the nodes, labeled with the object information. The edges are formed by directly-follows relationships between events for an object. 
%NEW WRITING
This graph describes the dependencies between events. Events with a path between them depend on each other, while event pairs with an absence of a path are independent. For example, $e_3$ is a prerequisite for $e_4$ and $e_5$. However, $e_4$ and $e_5$ have an arbitrary order. 
Since the object-centric event data exhibit a graph structure, we base our concept of a process execution on a graph rather than a sequence.
\iffalse
We use these notations to visualize the event-object graph. The nodes of this graph are events labeled with $obj_L$. The directed edges are given by the event pairs in $con_L$. An example constructed from the accompanying event log is depicted in \autoref{fig:event_data_example}.
The nodes are labeled with the objects of the events, e.g., $e_3$ is labeled with $obj_L(e_3)= \{o1,m1,m2\}$. We, furthermore, visualize the different types with different colors. Two events that follow each other in one of the object sequences are connected with an edge, e.g., $(e_1,e_3)\in con_L$ since $trace(m1) = \langle e_1,e_3,e_5,e_6\rangle$. 
The event-object graph defines the dependency among events, i.e., one event has to occur before the other. It furthermore defines which events do not depend on each other, i.e., the order of their execution is arbitrary. The existence of a path from one event to another one in the graph of the global partial order implies a dependence. For example, $e_3$ is a prerequisite for $e_4$ and $e_5$. However, $e_4$ and $e_5$ have an arbitrary order.  \fi
{\setlength{\belowcaptionskip}{-10pt}
\begin{table}
	\begin{minipage}{0.45\columnwidth}
	
\scriptsize{
    \centering
    \begin{tabular}{|c|c|c|c|}
    \hline
     \multirow{2}{*}{\textbf{Event}}& \multicolumn{1}{c|}{\textbf{Attr.}} & \multicolumn{2}{c|}{\textbf{Object Types}} \\ \cline{2-4}
          & $\ldots$ & Type1 & Type2  \\\hline\hline
         $e_1$   & $\ldots$& & m1 \\\hline  
         $e_2$  & $\ldots$ & & m2 \\\hline  
         $e_3$   & $\ldots$& o1 & m1, m2 \\\hline 
         $e_4$  & $\ldots$ & o1 &  \\\hline 
         $e_5$  & $\ldots$ &  & m1, m2 \\\hline 
         $e_6$  & $\ldots$ & o1  & m1, m2 \\\hline 
         $e_7$   & $\ldots$& o2 & m3, m4 \\\hline 
         $e_8$   & $\ldots$& o2 &  \\\hline 
         $e_9$   & $\ldots$& & m3 \\\hline  
         $e_{10}$  & $\ldots$ & & m4 \\\hline 
         $e_{11}$   & $\ldots$&  & m3, m4 \\\hline  
         $e_{12}$  & $\ldots$ & o2  & m3, m4 \\\hline 
    \end{tabular}
    
    \label{tab:ocel}
    }
	\end{minipage}\hfill
	\begin{minipage}{0.48\columnwidth}
		\centering
		\includegraphics[width=1\columnwidth]{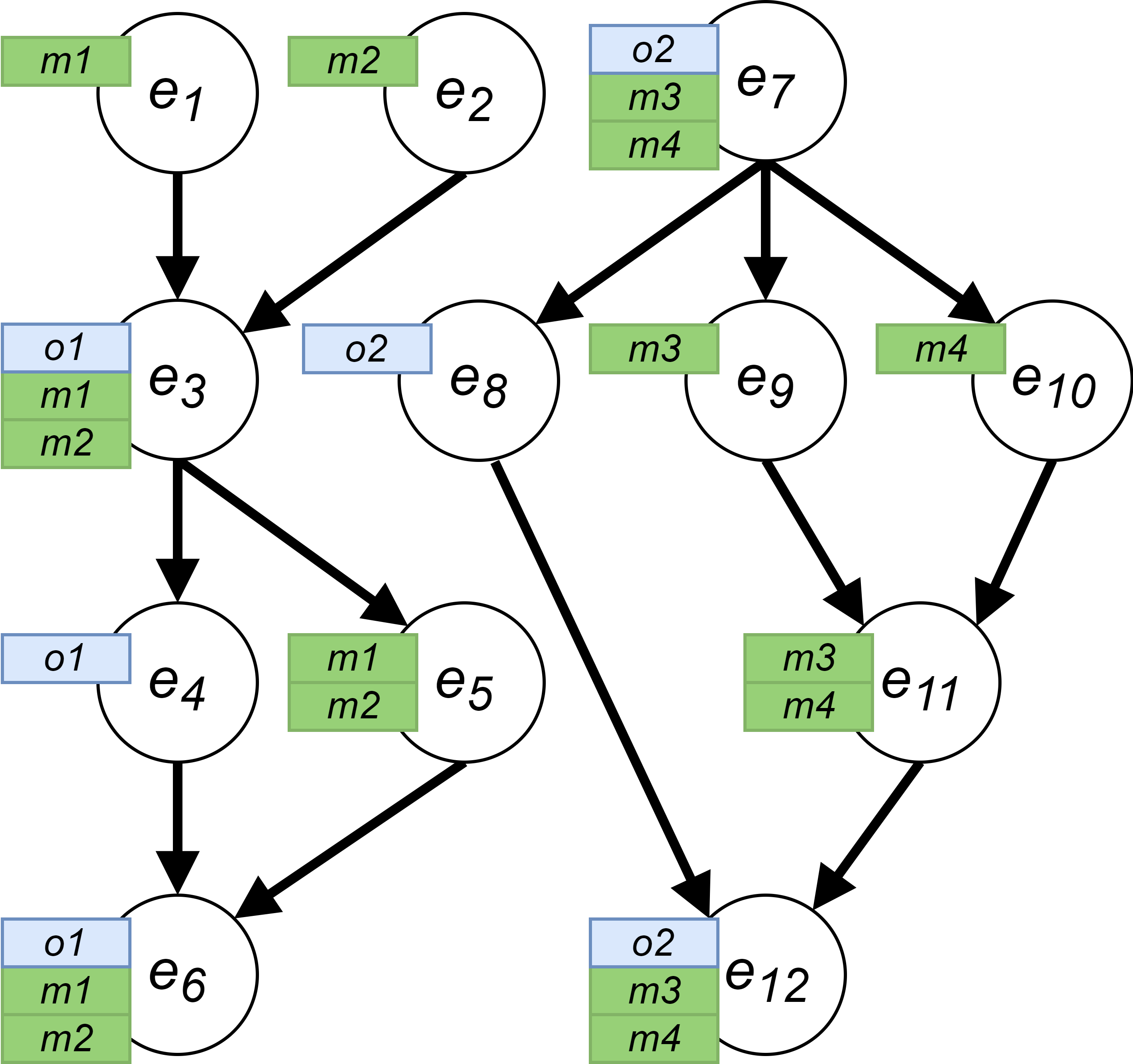}

	\end{minipage}
	\captionsetup{type=figure}
	\captionof{figure}{Object-centric event log and the corresponding event-object graph. Events following each other for one object are connected.}
	\label{fig:event_data_example}
\end{table}}
{\setlength{\belowcaptionskip}{-9pt}\begin{figure}[t]
    \centering
    \includegraphics[width = 0.4\columnwidth]{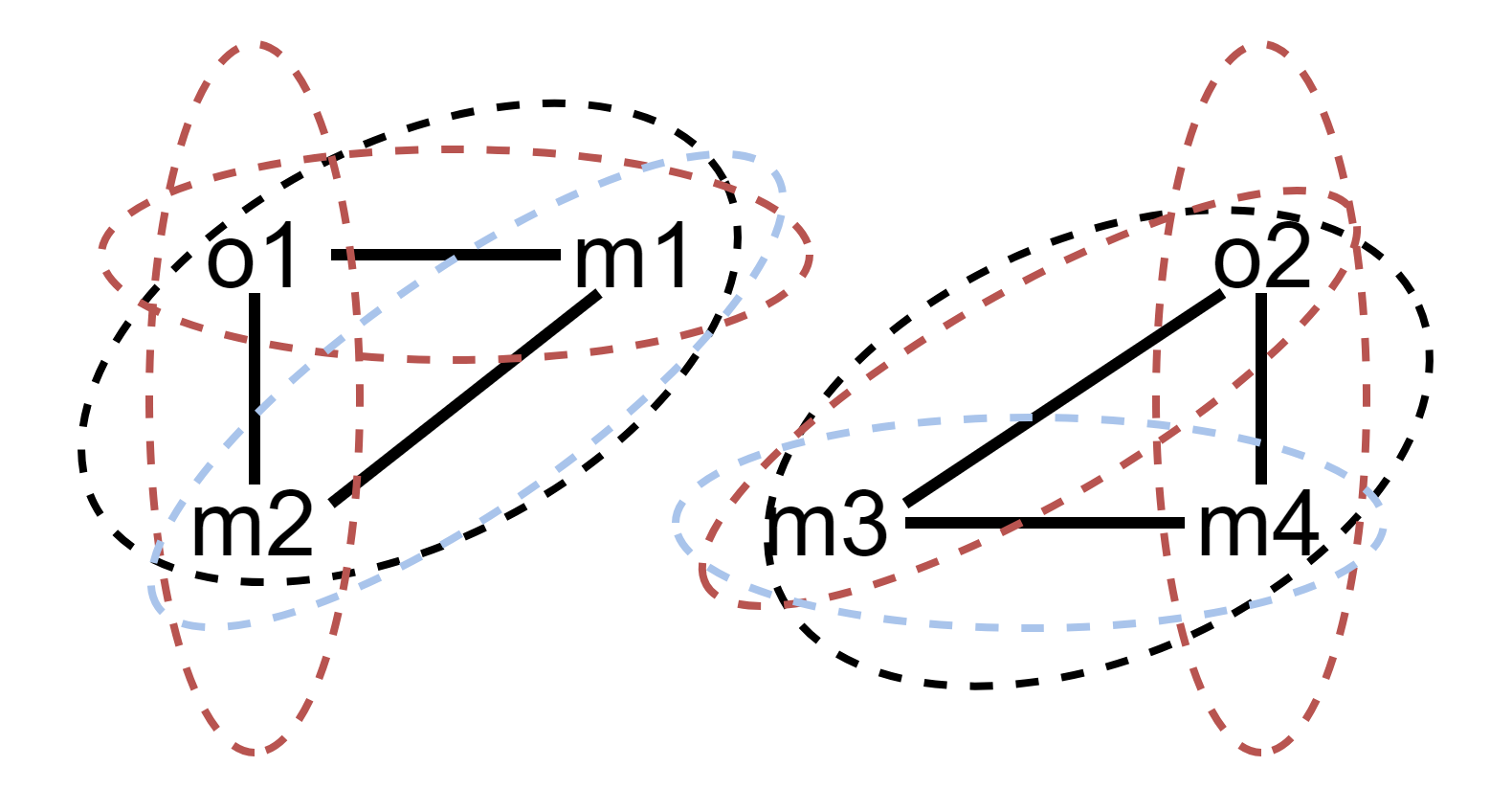}
    \caption{Example of the object graph for \autoref{fig:event_data_example}. Each connected subgraph, indicated by the dashed lines, can extract a process execution.}
    \label{fig:object_graph}
\end{figure}}
\section{Process Execution Extraction}
\label{sec:execandequiv}
This section introduces the concept of process executions and techniques to extract process executions from an object-centric event log. The case concept in traditional process mining describes the event sequence for a single object. We generalize this concept to process executions describing the event graphs of multiple interdependent objects. 
\subsection{Process Executions}
\label{sec:execution_extraction}
Before introducing the concept of a process execution, we first define the relationships between objects through the object graph. In this graph, objects form the nodes and are connected if they share an event. 
\begin{definition}[Object Graph]
Let $L{=}(E,O,\mathit{trace},\mathit{attr})$ be an event log. 
The object graph is an undirected graph $\mathit{OG}_L = (O,C_O)$ with $C_O {=} \{\{o1,o2\} {\subseteq} O \mid \exists_{e {\in} E} \ o_1,o_2{\in} obj_L(e) \wedge o_1 {\neq} o_2\}$. We denote the length of the shortest path between to objects $o,o'{\in} O'$ in $\mathit{OG}_L$ with $\mathit{dist}: O'\times O'\rightarrow \mathbb{R}\cup \{\perp \}$ where $\perp$ denotes the absence of a path.
\end{definition}
\autoref{fig:object_graph} depicts an example of the object graph for the object-centric event log in \autoref{fig:event_data_example}. The graph tells us which objects co-appear in events and, therefore, which objects depend on each other, directly and transitively. %\footnote{A traditional event log would produce a special case of an object graph: Since no event has more than one object, no edges would be introduced in the object graph.}

We now generalize the case concept used for traditional event data such  that a process execution from an object-centric event log covers multiple objects instead of one. However, these objects must be co-dependent, i.e., they must form a connected subgraph in the object graph. A process execution is a graph formed from the events and their directly-follows relationships for a set of connected objects.
\begin{definition}[Process Execution]
Let $L{=}(E,O, \mathit{trace}, \allowbreak \mathit{attr})$ be an event log and $O'\subseteq O$ be a subset of objects that forms a connected subgraph in $\mathit{OG}_L$.
The process execution of $O'$ is a directed graph $p_{O'}{=}(E',D)$ where 
\begin{itemize}
    \item $E' = \{e\in E \mid O' \cap \mathit{obj}_L(e) \neq \emptyset\}$ are the nodes, and
    \item $D = con_L\cap (E' \times E')$ are the edges.
\end{itemize} 
\end{definition}
The dashed lines in \autoref{fig:object_graph} indicate the different connected subgraphs of the object graph that each may define a process execution. As an example we focus on the black dashed lines, i.e., the two connected components of the object graph. Using each of these two object sets, we retrieve two process executions, which are equal two the weakly connected components of the event-object graph in \autoref{fig:event_data_example}.
%The retrieval of process executions from an object-centric event log is not predetermined. 
A selection of different subgraphs leads to an extraction of different process executions. This is by design, as object graphs and their dependencies can grow very large. Splitting this graph apart or selecting only some relationships of interest can be done by only selecting specific subgraphs. In the following section, we introduce two specific extraction techniques.
\paragraph{Process Executions Through Connected Components}
Our first technique retrieves the largest possible process executions, capturing all the dependencies and interactions contained in the event data. We do this by determining the connected components of the object graph. Each connected component is used to extract one process execution. 
The connected components of the object graph in \autoref{fig:object_graph} are indicated with black dashes.
\begin{definition}[Connected Component Extraction]
Let $L=(E,O,\mathit{trace},\mathit{attr})$ be an event log and $\mathit{OG}_L = (O,C_O)$ its object graph. $\mathit{ext}_\mathit{comp}(L) = \{p_{O'} \mid O' \subseteq O \wedge (O', C_O\cap (O' \times O') ) \text{ is a connected component of } \mathit{OG}_L\}$ extracts process executions by connected components.
\end{definition}
\iffalse Both of the weakly connected components of the global partial order of events in \autoref{fig:event_data_example} corresponds to one of the two connected components in the object graph. \fi
This technique has two main advantages. It is parameter-free, i.e., no interaction of the user is required, and it captures all dependencies between events and objects, forming a base for extensive exploration of interdependencies in the resulting process executions. However, this technique also comes with disadvantages. Connected components in the object graph can grow extremely big. In the worst case, the whole object graph is one single component leading to the extraction of only one single process execution. 
%Process executions grow increasingly big for increasingly large components, worsening accessibility and understandability.
For these reasons, we introduce a second technique of extracting process executions that neglects some connections to retrieve smaller process executions.
\paragraph{Process Executions Through a Leading Type}
Since every connected subgraph of the object graph can be used to extract process executions, we introduce a technique that identifies subgraphs of the object graph that revolve around a leading type. Such subgraphs have one node of an object of the leading type that is connected to nodes of objects of other types only by paths of the same length for each type.
To find these subgraphs, the object graph is traversed in a breadth-first manner for each object of the leading type. Objects for which objects of the same type were not already traversed in an earlier level are added to the process execution and to the breadth-first search. Other objects are discarded.
\begin{definition}[Leading Type Extraction]
Let $L{=}(E,O,\mathit{trace},\allowbreak \mathit{attr})$ be an event log, let $\mathit{OG}_L {=} (O,C_O)$ be its object graph and $\mathit{ot}{\in} \mathcal{T}$ be an object type.
$\mathit{ext}_{lead}(L, \mathit{ot}) {=} \{p_{O'} \mid o {\in} O \wedge \pi_\mathit{type}(o) {=} \mathit{ot} \wedge O' {=} \{o' {\in} O \mid \mathit{dist}(o,o'){\neq} {\perp} \wedge \neg \exists_{o'' \in O}\ \pi_\mathit{type}(o') {=} \pi_\mathit{type}(o'') \wedge  \mathit{dist}(o,o''){<}\mathit{dist}(o,o')\}$ retrieves process execution through the leading type $\mathit{ot}$.
\end{definition}
The idea behind these subgraphs is the following: For any given object, the objects that are closest in the object graph are assumed to be the ones with the most dependencies. These close objects should be added to a process execution until objects of the same type have already been added with a shorter path length. In this way, the closest objects of each (reachable) type are added.
In \autoref{fig:object_graph}, the subgraphs obtained with the leading type technique for type Type1 are indicated with black dashes, for type Type2 with red dashes. The light blue dashed lines indicate other possible subgraphs that could be used to extract process executions but are retrievable with neither of the two introduced techniques.
\section{Object-Centric Variants}
\label{sec:variants}
In general, determining similar process executions is a form of clustering. Some general function $f_{dist}(p,p')$ can be used to determine a distance between two process executions $p,p'$. Based on these distances, clustering of process executions can be performed. When speaking of equivalent process executions we deal with a specific instance of this problem. 
Two process executions can be equivalent with respect to an event attribute $a{\in} A$ if one cannot distinguish between the executions under consideration of the event's associated object types and the event's chosen attribute. If the considered attribute is the event's activity, one typically speaks of control-flow variants.
An $f_{dist}$ is necessary that yields a value of zero if and only if process executions are equivalent. Process executions with distance zero can subsequently be grouped into equivalence classes. The problem of determining graph isomorphism fulfills these criteria.

\subsection{Equivalence Class Mining}

First, we boil a process execution down to type, order, and attribute information by projecting it onto the event attribute and types of the events.
\begin{definition}[Projected Process Execution]
Let $L=(E,O,\mathit{trace},\allowbreak \mathit{attr})$ be an event log, $O'\subseteq O$ be a set of objects forming a connected subgraph in $\mathit{OG}_L$ and $p_{O'}=(E',D)$ the corresponding process execution. For an attribute $a\in \mathcal{A}$, the projected process execution $p_{O'}{\downarrow_a} = (E',D,l_e,l_d)$ is defined as a graph with two label functions:
\begin{itemize}
    \item $l_e(e) = (\pi_{attr}(e,a), \{(t,n) \in \mathcal{T} \times \mathbb{N}_0  \mid n =|\{ o {\in} O' \mid o {\in} obj_L(e) \wedge \pi_{type}(o) {=} t\}| \} )$ for $e \in E'$,
    \item $l_d((e_1,e_2)) = \{(t,n) \in \mathcal{T}\times \mathbb{N}_0  \mid n = |\{o {\in} O' \mid o \in obj_L(e_1) \wedge o {\in} obj_L(e_2) \wedge \pi_{type}(o) {=} t\}| \}$ for $(e_1,e_2) \in D$.
\end{itemize}
\iffalse
$P=(O_P,E_P)$ be a process execution of this event log with the associated graph $G_P = (E_P, D_P)$, and $a\in A$ be an event attribute that is defined for all events. The projected process execution $G_P{\downarrow_a} = (E_P,D_P,l_e,l_d)$ is a graph of nodes $E_P$, edges $D_P$ and labeling functions $l_e$ and $l_d$.  $l_e$ maps nodes to their event attribute values and number of objects of different types. The function $l_d$ maps edges to the number of objects of different types that induce these edges. The labeling functions are defined as follows:
\begin{itemize}
    \item $l_e(e) = (\pi_{attr}(e,a), \{(type,n) \in \mathcal{T} \times \mathbb{N}_0  \mid n =|\{ o {\in} O_P \mid o {\in} obj_L(e) \wedge \pi_{type}(o) {=} type\}| \} )$ for $e \in E_P$ labeling events with the attribute and the number of objets of different types,
    \item $l_d((e_1,e_2)) = \{(type,n) \in \mathcal{T}\times \mathbb{N}_0  \mid n = |\{o {\in} O_P \mid o \in obj_L(e_1) \wedge o {\in} obj_L(e_2) \wedge \pi_{type}(o) {=} type\}| \}$ for $(e_1,e_2) \in D_P$ labeling edges with the number of objects of different types they are induced by.
\end{itemize}\fi
\end{definition}
Two process executions $p,p'$ that are equivalent with respect to an attribute $a{\in} A$ if $p{\downarrow_a}$ and ${p'}{\downarrow_a}$ are isomorphic under consideration of the node and edge labels. 

Graph isomorphism is a well-studied problem in computer science for which, up to this point, no general polynomial-time algorithm is known \cite{graphisomorphism}. For the problem of determining groups of isomorphic graphs from a set of graphs, the naive solution would be to conduct a one-to-one isomorphism matching for each pair of graphs. We, however, employ a technique introduced by Rensink \cite{groove_twostep} in the \textsc{GROOVE} framework. This two-step technique first calculates an invariant hash code for each graph. Graphs with different hash codes can not be isomorphic. However, this does not mean graphs with the same hash code are isomorphic. Therefore, the initial equivalence classes of graphs with equal hash codes are refined through one-to-one comparisons. An additional benefit of employing this idea is retrieving an approximation of the equivalence classes by the initial, unrefined classes of the hash codes. The Weisfeiler-Lehman graph hashing \cite{graphhash} is used as a hashing function. We use the \textsc{VF2} algorithm \cite{vf2} for refinement of the initial classes, as it has shown superior performance for small, sparse graphs \cite{nautybadperformance}. We verify the results by conducting a one-to-one comparison with the \textsc{VF2} algorithm. As the calculation of graph hashes is known to have scalability problems for some graphs \cite{exponentialcanoncial}, we compare the running times of the employed two-step technique with a one-to-one checking using the \textsc{VF2} algorithm. \iffalse\autoref{alg:ALG3} depicts the our employed two-step algorithm.\fi
\iffalse
\begin{algorithm}
        \caption{Equivalence Class Calculation Through Graph Isomorphism}
        \label{alg:ALG3}
        \vspace{-4mm}
        \hrulefill
        
        \begin{algorithmic}
        \Require $P$ - list of projected process executions
        \Ensure $EQ$ list of lists with equivalent process executions
        \State hash\_map $\gets$ weisfeiler-lehman hash for each $p{\in}P$
        \State $EQ \gets []$ 
        \State $EQ' \gets $ \{hash: list of $p{\in}P$ with this hash\}
        \For{$eq' \in$ keys of $EQ'$}\Comment{Refine classes}
        \State refined\_$eq' \gets []$
        \For{$exec$ in $EQ'[eq']$}
        \For{$eq''{\in}$ keys of refined\_$eq'$}
        \If{$exec$ isomorph to refined\_$eq'[eq''][0]$}
        \State refined\_$eq'[$hash\_map[exec]$] \gets_{append} exec$
        \Else
        \State   refined\_$eq'[$hash\_map[exec]$] \gets [exec]$
        \EndIf
        \EndFor
        \EndFor
        \State $EQ \gets_{append}$  refined\_$eq'$
        \EndFor
        \end{algorithmic}
    \end{algorithm}\fi

We denote this two-step function to derive equivalence classes by $\mathit{iso}_a{:} \{p_1,\ldots, p_n\}{\rightarrow} \{1,\ldots,m\}$, retrieving $m{\leq} n$ equivalence classes for process executions $\{p_1,\ldots, p_n\}$ and an attribute $a\in A$. An equivalence class is a set of process executions $EQ_a {=} \{p_1,\ldots,p_k\}$ such that $\mathit{iso}_a(p_1) {=} \ldots {=}  \mathit{iso}_a(p_k)$. The set of all mined equivalence classes is denoted by $\mathcal{EQ}_a$. The relative frequency of an equivalence class $EQ_a {\in} \mathcal{EQ}_a$ is given by
$$f_{EQ_a} = \frac{|EQ_a|}{\sum_{EQ_a' \in \mathcal{EQ}_a} |EQ_a'|}.$$
\subsection{Variant Visualization}
\label{sec:Visualization}
{\begin{table}[t]
    \centering
    \caption{Example of the visualization of two variants for ordering machines using the activity attribute.\iffalse Events can be of type order and/or machine and the attribute chosen for the projected process executions is the activity of an event, i.e., what action is conducted.\fi}
    \footnotesize{
    \begin{tabular}{|m{0.55cm}|m{6.9cm}|}\hline
    \multicolumn{1}{|l|}{Class} &  Equivalence Class Visualization \\\hline

        1   & \vspace{0.08cm} \includegraphics[clip, trim=0cm 18.1cm 40.2cm 0cm, width=0.80\columnwidth]{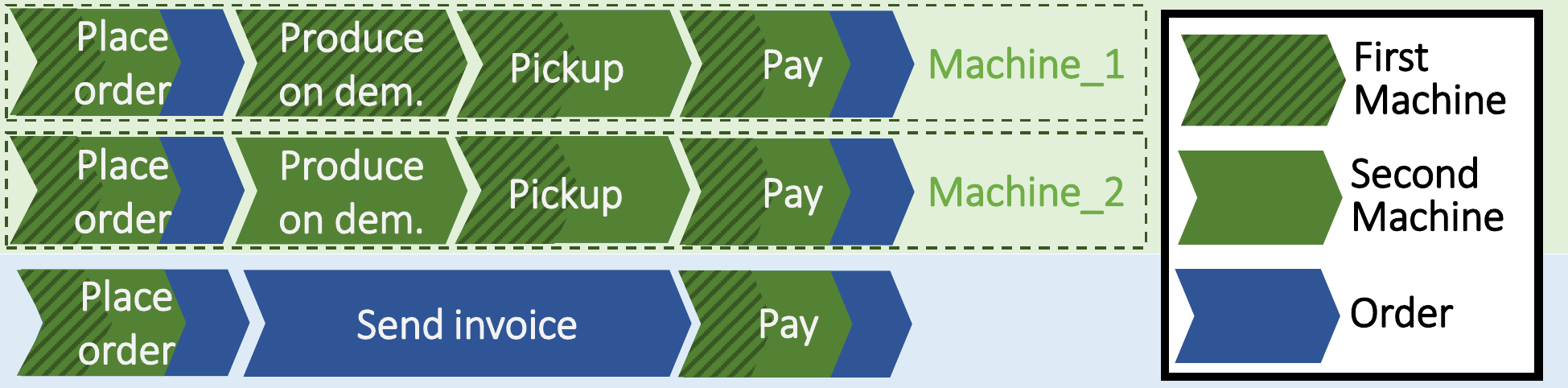} \\\hline
        2   & \vspace{0.08cm} \includegraphics[clip, trim=0cm 13.0cm 40.2cm 5.0cm, width=0.80\columnwidth]{Visualization_Example.pdf}  \\\hline
    \end{tabular}
    }
    \label{tab:ExampleVariants}
\end{table}}
{\footnotesize
\setlength{\tabcolsep}{0.5em}
\begin{table*}[t]
    \centering
    \caption{Overview of extracted process executions for different techniques, their properties, and variants.}
    \label{tab:results}
    \resizebox{0.99\textwidth}{!}{
    \begin{tabular}{|l|c|c|c|l|c|c|c|c|}\hline
    \multirow{2}{*}{Data Set} & Number of & Number of & Number of &  \multirow{2}{*}{Extraction Technique} & Number of & Events per Execution & Objects per Execution & Number of\\
     & Events & Types & Objects &  & Executions &  (max, min, avg) & (max, min, avg)  & Variants\\\hline
     \multirow{3}{*}{$\boldsymbol{DS_1}$ Loan Application Process} &  \multirow{3}{*}{$507553$} &  \multirow{3}{*}{$2$}  &  \multirow{3}{*}{$67498$} & Connected components & 28509 & $(61,7,17.8)$ & $(11,2,2.4)$ & $3420$ \\\cline{5-9}
       &&&& Leading type: Application &28509 & $(61,7,17.8)$ & $(11,2,2.4)$ & $ 3420$ \\\cline{5-9}
       &&&& Leading type: Offer &38989 & $(56,7,18.7)$ & $(2,2,2)$ & $7763 $ \\\hline
        \multirow{4}{*}{$\boldsymbol{DS_2}$ Order Management Process} &\multirow{4}{*}{$22367$} &\multirow{4}{*}{$3$} &\multirow{4}{*}{$11484$} & Connected components & 83 & $(1382,8,269.5)$ & $(717,3,138.4)$ & $83$ \\\cline{5-9}
       &&&& Leading type: Items & 8159 & $(155,8,57.9)$ & $(11,3,6)$ & 8155\\\cline{5-9}
       &&&& Leading type: Orders & 2000 & $(179,8,51.3)$ & $(68,3,19.4)$ &1998\\\cline{5-9}
       &&&& Leading type: Packages & 1325 & $(155,8,49.7)$ & $(32,3,10.5)$ &1325\\\hline
        \multirow{3}{*}{$\boldsymbol{DS_3}$ Customer Incident Management} & \multirow{3}{*}{$119998$} & \multirow{3}{*}{$2$} & \multirow{3}{*}{$25598$} & Connected components & 4825 & $(259,2,24.3)$& $(53,2,5.1)$ & 3388\\\cline{5-9}
       &&&& Leading type: Incident & 19966 & $(144,2,25.2)$& $(3,2,2)$ & 16935\\\cline{5-9}
       &&&& Leading type: Customer & 4826 & $(259,2,24.3)$& $(53,2,5.1)$ & 3389 \\\hline
       \multirow{7}{*}{$\boldsymbol{DS_4}$ Farmer Subsidies Process}  & \multirow{7}{*}{$852610$} & \multirow{7}{*}{$6$}& \multirow{7}{*}{$58747$} & Connected components & 14507 & $(2973,31,58.8)$ & $(22,3,4.2)$ & 7274 \\\cline{5-9}
       &&&& Leading type: Payment Application & 14507 & $(2973,31,58.8)$ & $(22,3,4.2)$ & 7274 \\\cline{5-9}
       &&&& Leading type: Control summary & 14507 & $(2973,31,58.8)$ & $(22,3,4.2)$ & 7274 \\\cline{5-9}
       &&&& Leading type: Entitlement application & 205 & $(279,43,68.6)$ & $(9,5,5.1)$ & 190 \\\cline{5-9}
       &&&& Leading type: Geo parcel document  & 14507 & $(2973,31,58.8)$ & $(22,3,4.2)$ & 7274 \\\cline{5-9}
       &&&& Leading type: Inspection  & 1999 & $(2941,48,155.2)$ & $(6,5,5)$ & 1955\\\cline{5-9}
       &&&& Leading type: Reference Alignment & 14503 & $(2973,31,58.8)$ & $(22,4,4.2)$ &7270\\\hline
         
    \end{tabular}}
\end{table*}}
\begin{algorithm}
        \caption{Horizontal positioning of events}
        \label{alg:ALG2}
        %\vspace{-4mm}
        %\hrulefill
        \begin{algorithmic}
        \Require $p_{O'}{\downarrow}_a$ - projected process execution, $e{\in}\mathcal{E}$ - event
        \Ensure $x_{start}, x_{end}$ horizontal position for the event
        \State $x_{start} \gets $  \Call{get\_x\_start}{$p_{O'}{\downarrow}_a, e$}
        \State $x_{end} \gets$ \Call{get\_x\_end}{$p_{O'}{\downarrow}_a, e$}
        \Function{get\_x\_start}{$p_{O'}{\downarrow}_a, e$}
        \State pre $\gets$ predecessors of $e$ in $p_{O'}{\downarrow}_a$
        \If{$|pre| = 0$}
        \State return x=0
        \EndIf
        \State return max(\Call{get\_x\_start}{$p_{O'}{\downarrow}_a, e'$} for e' in pre)$+1$
        \EndFunction
        \Function{get\_x\_end}{$p_{O'}{\downarrow}_a, e$}
        \State suc $\gets$ successors of $e$ in $p_{O'}{\downarrow}_a$
        \If{$|suc| = 0$}
        \State return \Call{get\_x\_start}{$p_{O'}{\downarrow}_a, e'$}
        \EndIf
        \State return min(\Call{get\_x\_start}{$p_{O'}{\downarrow}_a, e'$} for e' in suc)$-1$
        \EndFunction
        \end{algorithmic}
    \end{algorithm}
    
The techniques introduced before allow us to determine frequent process executions w.r.t. an event attribute. However, the mathematical object retrieved by these techniques is a graph with complex edge and node labels. These graphs are hard to interpret and understand, especially with increasing size. Therefore, we focus on delivering an improved visualization for control-flow variants using the activity attribute.
To provide an accessible and understandable visualization, we introduce an extension of variant visualization used in process mining and business process management \cite{BPM,Cortado} for control-flow variants. Traditionally, variants are visualized as a sequence of chevrons containing the activity labels.

An example of the result from the proposed visualization technique is depicted in \autoref{tab:ExampleVariants} for two variants.   
There are two object types: orders (blue) and machines (green). Both variants have two objects of type machine, indicated by different shades of green. In the first variant, an order is placed for two machines. While an invoice for the order is sent, the machines are each produced and picked up concurrently, i.e., these events happen in an arbitrary order. For example, sending the invoice is not dependent on the production of any of the machines. To conclude, the order for both machines is paid. In the second variant, the machines are produced to stock in the beginning.

In our visualization, each type has a specific base color, and each object of such a type has a particular shade of the base color. Each object has a lane showing the event sequence for the object. Each event is depicted as a chevron with the activity label inside the chevron. If an event is shared between objects, it is placed on all corresponding object lanes and colored with the corresponding colors. The shared events are placed at the same horizontal position to respect the partial orders between events.
Since the horizontal position of events depends on the previously occurring shared events and their predecessors, we introduce a layouting algorithm to determine the horizontal positions of events. The objects determine the vertical position, each object is associated to one unique vertical position grouped by type. \autoref{alg:ALG2} describes retrieve the starting and ending horizontal position of an event. The starting position is determined recursively based on the predecessor events. The ending position is based on the starting position of the successor events.

\section{Algorithmic Evaluation}
\label{sec:evaluation}
{\setlength{\belowcaptionskip}{-10pt}\begin{figure}[t]
    \centering
    \includegraphics[width=\columnwidth]{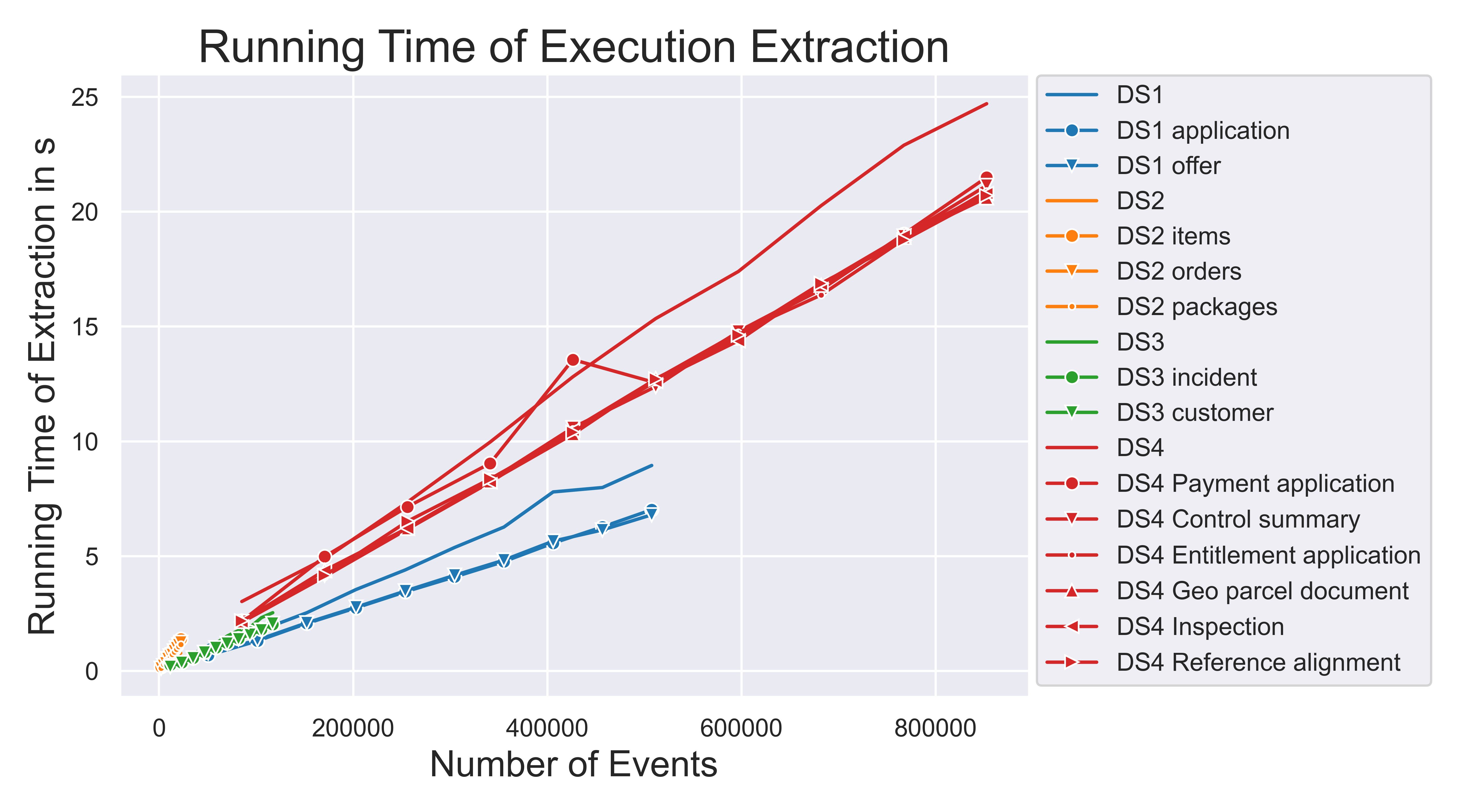}
    \caption{Running times of execution extraction for different subset sizes, logs, and extraction techniques.}
    \label{fig:runningtimes1}
\end{figure}}
In this section, we evaluate the four contributions proposed in this paper. We use four event logs described in \autoref{tab:results}. Three of them, $\boldsymbol{DS_1}$~\cite{BPI2017}, $\boldsymbol{DS_3}$ and $\boldsymbol{DS_4}$~\cite{BPI2018}, are real-life event logs, while $\boldsymbol{DS_2}$ is a synthetic event log, consisting of an especially high amount of connected objects and variability. First, we evaluate the results and running times of the different process execution extraction techniques. Then, we evaluate our employed two-step equivalence class calculation technique \cite{groove_twostep} and compare the running time to a one-to-one matching using the \textsc{VF2}-algorithm \cite{vf2}. We use the event activity attribute for each data set to determine equivalence classes, i.e., we calculate variants. 
%%%%%%%% MAYBE for an extension
\iffalse
Furthermore, we disassemble the running times and results from the two steps of calculating equivalence classes, i.e., determining initial classes and refining them. \fi
At last, we show the running times of our equivalence class layouting algorithm.
\begin{figure*}[t]
  \centering
  \subfloat[][$\boldsymbol{DS}_1$]{\includegraphics[width=.235\textwidth]{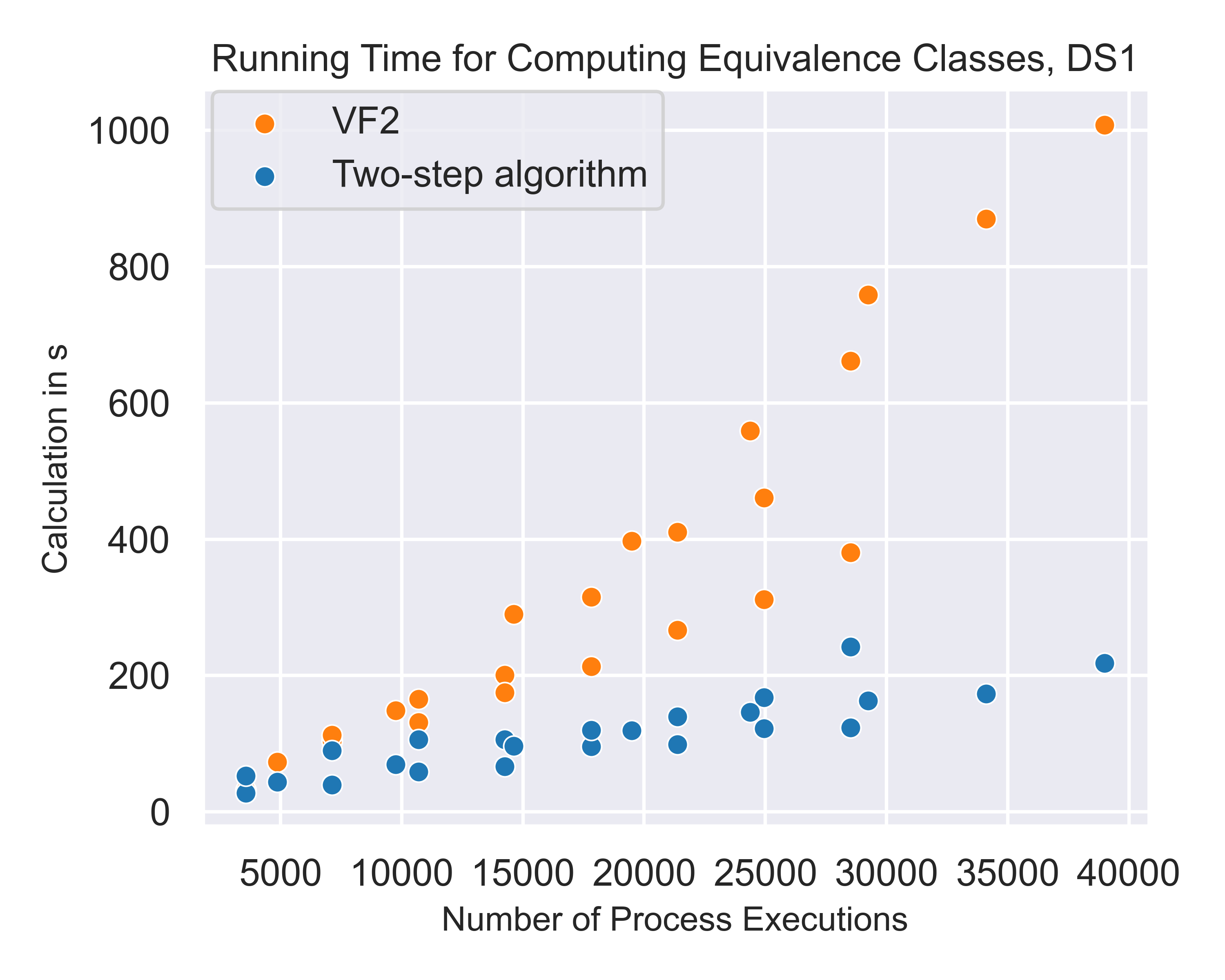}}\quad
  \subfloat[][$\boldsymbol{DS}_2$]{\includegraphics[width=.235\textwidth]{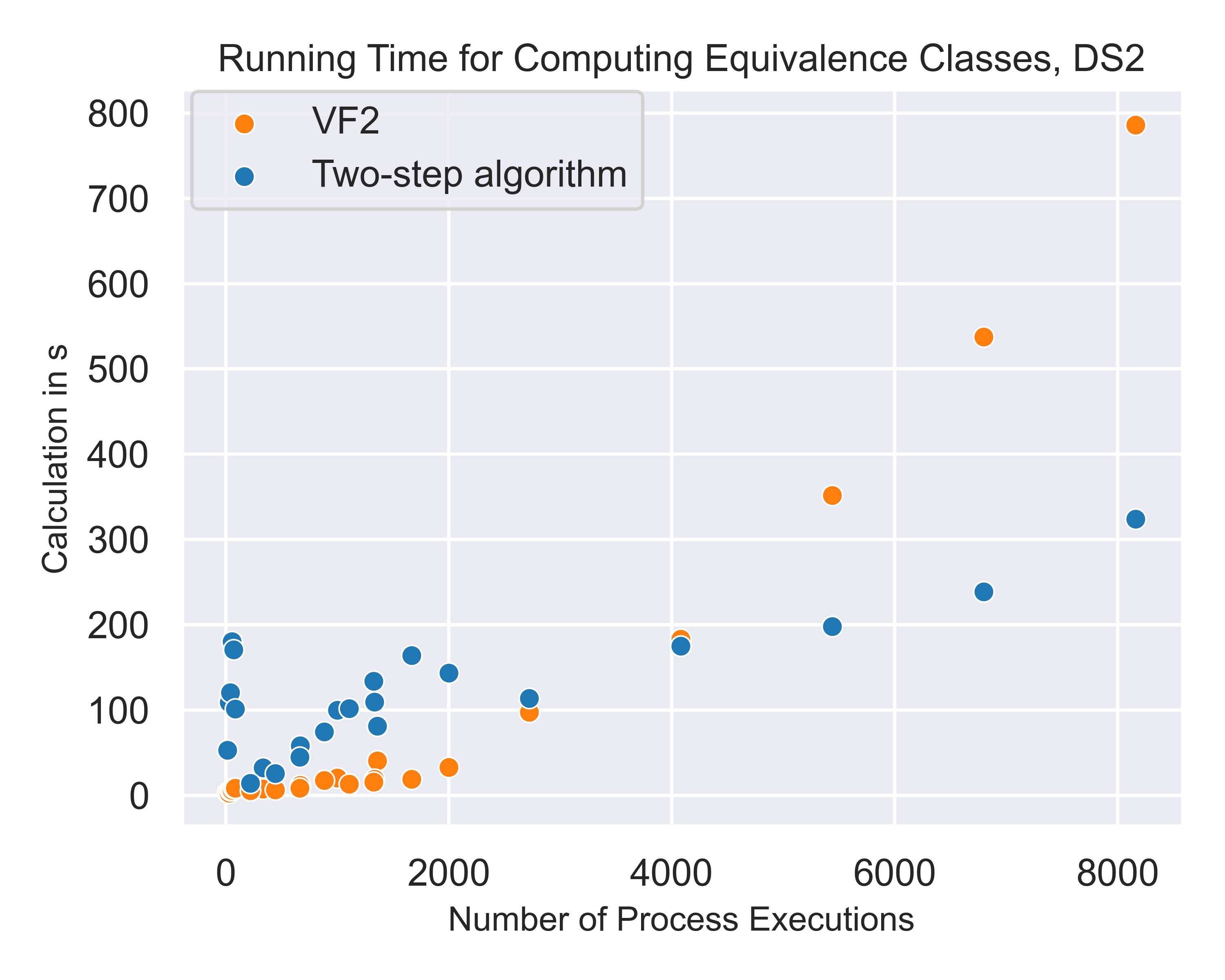}}\quad
  \subfloat[][$\boldsymbol{DS}_3$]{\includegraphics[width=.235\textwidth]{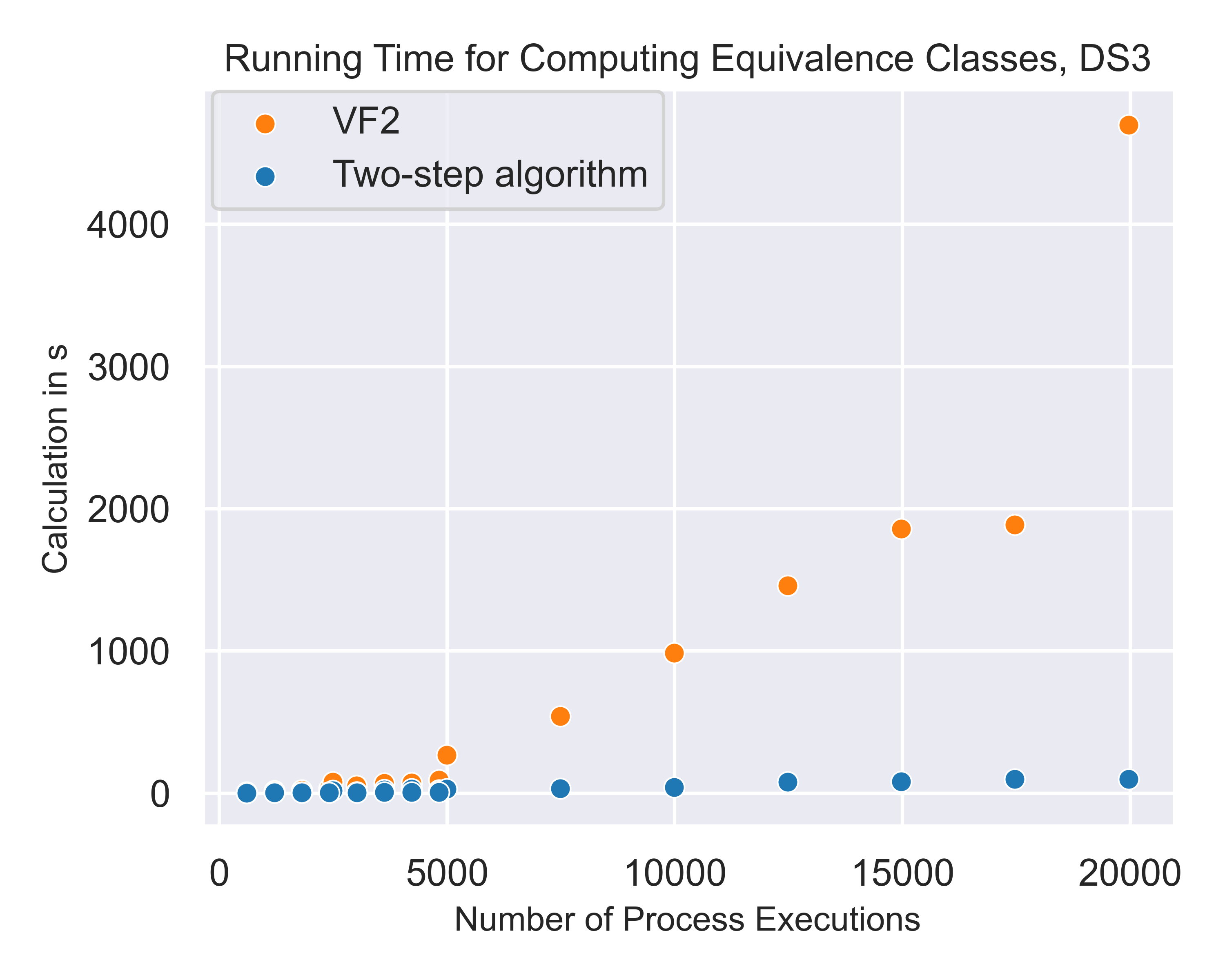}}\quad
  \subfloat[][$\boldsymbol{DS}_4$]{\includegraphics[width=.235\textwidth]{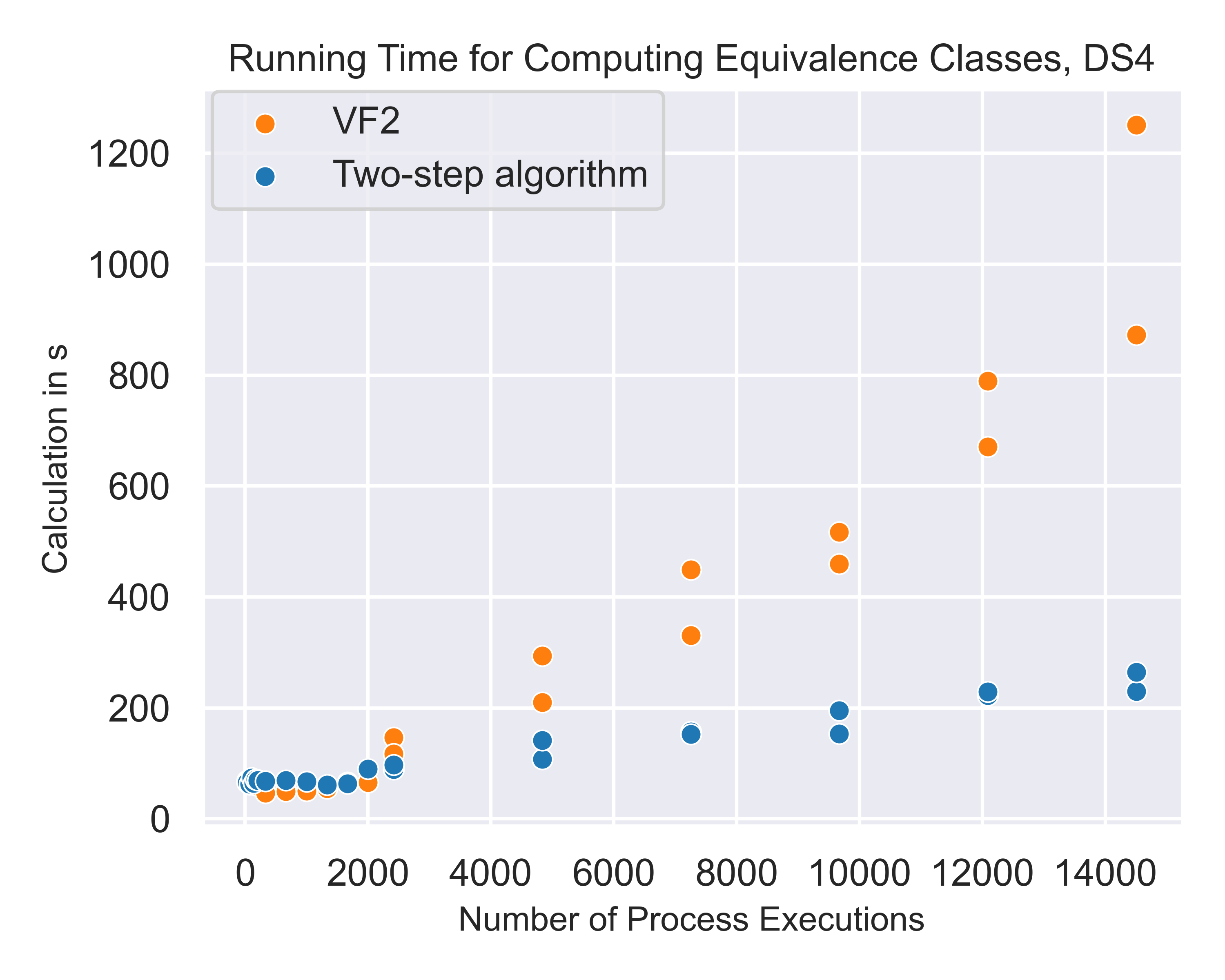}}
  \caption{Running times of the two-step algorithm compared to the \textsc{VF2}-algorithm for determining isomorphism.}
  \label{fig:runningtimes2}
\end{figure*}
Our experiments and data sets are publicly available on GitHub\footnote{\url{https://github.com/niklasadams/OCCasesAndVariants.git}}. 
The tool OC$\pi$~\cite{OCPI}\footnote{\url{https://www.ocpi.ai}} can be used to explore the individual variants of the different event logs in and end-to-end application of our contributions.
\iffalse
Furthermore, a reduced version of OC$\pi$~\cite{OCPI} preloaded with the evaluation's event logs is freely available\footnote{\url{https://drive.google.com/file/d/16EvJJDE_tPYcchXmd3aGBUbxjKZzjUDV/view?usp=sharing} Please run \textit{frex.exe}.}.
This tool contains the full end-to-end application of our contributions and enables the exploration of variants and their frequencies.\fi

\subsection{Process Execution Extraction}
\autoref{tab:results} depicts the number of executions, the maximum, minimum, and average number of events and objects per execution for all data sets, and the two introduced execution extraction techniques. For leading type, the extraction is performed for each type.
For some data sets, like $\boldsymbol{DS_1}$, the results are quite similar. However, for $\boldsymbol{DS_2}$ the raw data contains large connected components where many objects get entangled, visible by the low number of executions with connected components and their high average number of events. Using leading type, the executions increase in number but decrease in size, allowing for a tighter focus.

\autoref{fig:runningtimes1} shows the running times for each execution extraction technique for different sizes of each event log. Line plots without a marker refer to connected components extraction. A linear development of the running times can be observed for the given sublogs. A relationship of the slope and characteristics of the event logs, e.g., the average size of executions, seems likely. In general, the extraction by leading types is slightly faster than connected components. Both, connected components and leading type, show promising scalability for the application on real-life event data.
\iffalse
skeleton table
\begin{table*}[t]
    \centering
    \caption{Properties of executed process executions}
    \begin{tabular}{|l| c|c|   c|c| c|c| c|c|   c|c| c|c| c|c| }
    \hline
        \multirow{3}{*}{}&\multicolumn{14}{c|}{$\boldsymbol{DS_1}$}\\\cline{2-15}
        &\multicolumn{2}{c|}{\multirow{2}{*}{\#exec}}&\multicolumn{6}{c|}{\multirow{1}{*}{\#events/exec}}&\multicolumn{6}{c|}{\multirow{1}{*}{\#objs./exec}}\\\cline{4-15}
        &\multicolumn{2}{c|}{}&\multicolumn{2}{c|}{max}&\multicolumn{2}{c|}{min}&\multicolumn{2}{c|}{avg}&\multicolumn{2}{c|}{max}&\multicolumn{2}{c|}{min}&\multicolumn{2}{c|}{avg}\\\hline\hline
       
         Classic (one seq. $=$ one exec.)&\multicolumn{2}{c|}{} &\multicolumn{2}{c|}{} &\multicolumn{2}{c|}{}  &\multicolumn{2}{c|}{}  &\multicolumn{2}{c|}{}&\multicolumn{2}{c|}{}&\multicolumn{2}{c|}{}\\\hline
         Weakly Con. Comp.&\multicolumn{2}{c|}{} &\multicolumn{2}{c|}{} &\multicolumn{2}{c|}{}  &\multicolumn{2}{c|}{}  &\multicolumn{2}{c|}{}&\multicolumn{2}{c|}{}&\multicolumn{2}{c|}{}\\\hline
         Leading Type &&&& && &&&&&&&&\\\hline
    \end{tabular}
    \label{tab:event logs}
\end{table*}
\fi

\subsection{Variants}
%%%%%%%%%%5COULD BE PUT IN FOR AN EXTENSION
\iffalse
{\setlength{\belowcaptionskip}{-10pt}\begin{figure}[t]
    \centering
    \includegraphics[width = \columnwidth]{first_second_new_.png}
    \caption{Running times of both steps and the resulting number of equivalence classes compared.}
    \label{fig:twosteps}
\end{figure}}\fi
In this section, we first discuss the number of variants calculated for each event log. We then compare the results of our employed two-step technique with a baseline technique of determining isomorphic graphs from a set of graphs. 
%%%%%%%%%%%%%%%COULD BE PUT IN FOR AN EXTENSION
\iffalse
Furthermore, we show the running times and results of both algorithm steps, i.e., how much time each step consumes and how many equivalence classes are retrieved.\fi

For the real-life logs, i.e., $\boldsymbol{DS_1}$, $\boldsymbol{DS_3}$, and $ \boldsymbol{DS_4}$, the number of variants is often significantly smaller than the number of executions (cf. \autoref{tab:results}). This observation shows that these types of event data contain several equivalent process executions. Only for the synthetic data set $\boldsymbol{DS_2}$ the number of equivalence classes is almost the number of process executions. Through this event log's highly entangled and interconnected nature, almost no process execution is equivalent to any other execution. A clustering or subgraph mining approach might be more suited for such extreme cases.

We evaluate our employed technique's correctness and running time against a baseline. This baseline works similarly to the two-step technique. However, it starts with all executions in one equivalence class and then refines it. While doing so, it performs a matching under consideration of edge and node labels using the \textsc{VF2}-algorithm.

\autoref{fig:runningtimes2} depicts the running times for our employed technique and the baseline, i.e., \textsc{VF2}, for each event log. The measures are collected by calculating equivalence classes for different subset sizes of the process executions from different extraction techniques. Generally, the two-step approach shows good scalability and efficiency compared to the baseline technique.
\subsection{Variant Visualization}
In this section, we evaluate the scalability of our variant layouting algorithm. We perform the layouting for all executions retrieved for all logs by extracting connected components. We sort the running times for each equivalence class based on the number of events in a corresponding process execution and plot the results in \autoref{fig:layoutingrunning}. Furthermore, each point is colored and sized according to the number of objects associated with the equivalence class. We observe increasing running times with an increase of events and objects. In general, the results show promising scalability of the layouting. 
{\setlength{\belowcaptionskip}{-10pt}\begin{figure}[t]
    \centering
    \includegraphics[width=\columnwidth]{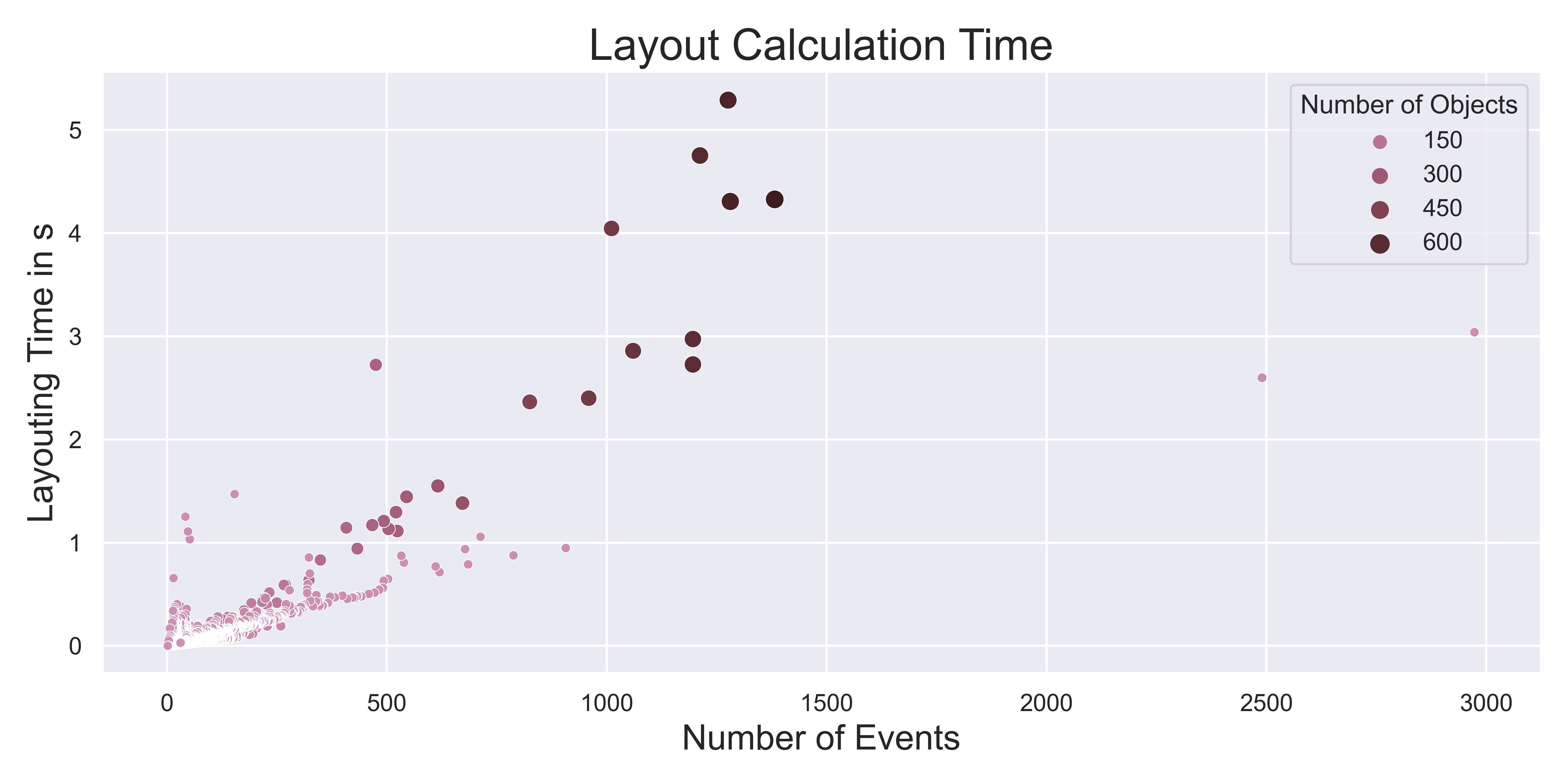}
    \caption{Running time of the layouting algorithm depending on the number of events and the number of objects, indicated by the size and color of the points. The running times are collected over all event logs and variants.}
    \label{fig:layoutingrunning}
\end{figure}}
\iffalse
{\scriptsize
\begin{table}[]
    \centering
    \caption{Running times disassembled}
    \begin{tabular}{|c|c|c|c|c|} \hline
         & total &execution extraction&equivalence classes & layouting \\ \hline
         $\boldsymbol{DS_1}$& & & &\\\hline
    \end{tabular}
    
    \label{tab:allrunningitmes}
\end{table}}\fi
\section{End-to-End Application}
\label{sec:case}
{
\begin{table*}[htp!]
    \centering
    \caption{The three most frequent object-centric variants in the loan application process. Additionally, we depict two variants with multiple offers for one application. Activities are abbreviated using the first letters.}
    \footnotesize{
    \begin{tabular}{|m{0.7cm}|m{15.7cm}|}\hline
    \multicolumn{2}{|p{16.4cm}|}{\textbf{Activity abbreviations:} ACA = Create Application, AS = Application Submitted, ACon = Application Concept, WCA = Application Workflow Completed, AA = Application Accepted, OCO = Create Offer, OCre = Offer Created, OSMO = Offer Sent Mail \& Online, WCAO = Calling After Offer, ACom = Application Complete, ACan = Application Cancelled, OCan = Offer Cancelled, OR = Offer returned, WVA = Validate Application, AV = Validating, WCIF = Calling for Incomplete Files, AI = Application Incomplete, OA = Offer Accepted, AP = Pending   }\\\hline
    \multicolumn{2}{|l|}{\textbf{Variant number} (corresponding frequency) \& visualized \textbf{object-centric variant} }\\\hline

        1 (11\%)  & \vspace{.1cm}\includegraphics[clip, trim=0cm 19.6cm 14cm 0cm, width=.9\columnwidth]{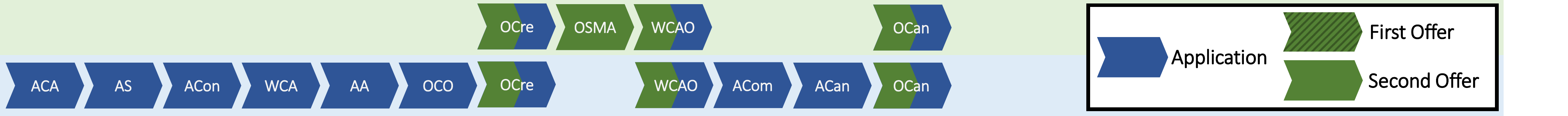} \\\hline
        2 (5\%)  & \vspace{.1cm}\includegraphics[clip, trim=0cm 16.3cm 14cm 3.40cm, width=.9\columnwidth]{Visualization_BPI.pdf}  \\\hline
        3 (4\%)  & \vspace{.1cm}\includegraphics[clip, trim=0cm 12.88cm 14cm 6.81cm, width=.9\columnwidth]{Visualization_BPI.pdf} \\\hline
        14 (2\%)  & \vspace{.1cm}\includegraphics[clip, trim=0cm 7.75cm 14cm 10.22cm, width=.9\columnwidth]{Visualization_BPI.pdf} \\\hline
        35 ($0.4\%$)  & \vspace{.1cm}\includegraphics[clip, trim=0cm 2.62cm 14cm 15.28cm, width=.9\columnwidth]{Visualization_BPI.pdf} \\\hline
    \end{tabular}
    }
    \label{tab:BPIVariants}
\end{table*}}
This section showcases the end-to-end application of our contributions to an event log to retrieve insights about the most frequent variants. We use the event log provided by $\boldsymbol{DS_1}$. This data set describes customers applying for loans in a financial institution. After assessing an application, a loan offer is made. This offer can be accepted, refused, or canceled. Subsequently, new offers for the same application can be made. We use connected components/leading type application (both lead to the same result) to extract process executions. We extract the object-centric variants and visualize them using our layouting algorithm. \autoref{tab:BPIVariants} depicts the results. 

11\%, which amounts to more than 3000 process executions, show equivalent behavior: After some application steps are performed, an offer is created and sent to the customer. After a phone call, the application and the offer are canceled. However, in the second most frequent execution, the offer is returned and accepted after the phone call. Executions with multiple offers are also relatively frequent. The fourth and fifth depicted equivalence classes show executions where the second offer is canceled or accepted. The variant visualization provides intuitive insights into the creation and concurrency of objects.

\section{Conclusion}
\label{sec:conclusion}
This paper presented four contributions to translate the concept of cases and variants to object-centric event data. The case concept is generalized to process executions which are event graphs of multiple, dependent objects. We use connected subgraphs of the object graph to extract process executions and provide two specific algorithms. Using graph isomorphism algorithms we can determine equivalent process executions. We use these algorithms to determine object-centric variants and propose a visualization extending traditional variant visualization. In \autoref{sec:case}, we provided an end-to-end application of all these contributions to visualize the object-centric variants of a loan application process.\iffalse  First, we provided a general definition of process executions and a general mechanism to extract them. Furthermore, we defined two specific extraction techniques. Second, we proposed the employment of a two-step algorithm to determine equivalent process executions based on graph isomorphism. The first step may be sufficient to retrieve approximated equivalence classes in reduced time. Third, we proposed a scalable layouting algorithm that is used to create a visualization of equivalence classes. In \autoref{sec:case}, we apply these three contributions to show insights into the frequent executions of a loan application process. Furthermore, we provide a software tool to explore frequent process executions of the data sets used in this paper. 
The techniques and definitions proposed in this paper can be used to extract process executions from heterogeneous event data with shared events, building a foundation for future analysis and prediction techniques.\fi

%\paragraph{Future Work}
The work presented in this paper can be extended in two major directions: Additional extraction techniques and different clustering techniques. The two extraction techniques introduced are just two of many possible ones that could extract patterns of interest from the object graph. Our equivalence class calculation is one specific type of clustering. Process executions could be clustered in other ways based on other distance measures. Clustering would also help with generating insights for highly entangled event sequences. 
\iffalse
Furthermore, an extensive investigation of automorphism calculation can be conducted, considering multiple automorphism checking techniques, like \textsc{Nauty} \cite{practicalgraphisomorphism}, and an automatic choice of the technique based on properties of the process execution graph.\fi
\iffalse
Other research directions include creating projected process executions for partially available attributes, and mining local patterns across process executions.\fi

Cases are essential for traditional process mining techniques. With this paper, we provide a technique to extract the object-centric equivalent of cases. These can be used to adapt existing process mining techniques to the object-centric setting and develop new techniques providing novel insights.
\iffalse
\section*{Acknowledgment}

The preferred spelling of the word ``acknowledgment'' in America is without 
an ``e'' after the ``g''. Avoid the stilted expression ``one of us (R. B. 
G.) thanks $\ldots$''. Instead, try ``R. B. G. thanks$\ldots$''. Put sponsor 
acknowledgments in the unnumbered footnote on the first page.
\fi

\bibliographystyle{IEEEtran}

\bibliography{bibliography}

\end{document}